\documentclass[twocolumn]{aastex63}

\graphicspath{{./}{figures/}}
\usepackage{pifont}
\usepackage{multirow}
\usepackage{amsmath}
\usepackage{color}
\usepackage{graphicx}
\usepackage{subfigure}
\usepackage{boxedminipage}
\usepackage{threeparttable}

\begin{document}

\title{Long-duration Gamma-ray Burst and Associated Kilonova Emission from Fast-spinning Black Hole--Neutron Star Mergers}

\author[0000-0002-9195-4904]{Jin-Ping Zhu}
\affiliation{Department of Astronomy, School of Physics, Peking University, Beijing 100871, China; \url{zhujp@pku.edu.cn}}

\author[0000-0002-9738-1238]{Xiangyu Ivy Wang}
\affiliation{School of Astronomy and Space Science, Nanjing University, Nanjing 210093, China}
\affiliation{Key Laboratory of Modern Astronomy and Astrophysics (Nanjing University), Ministry of Education, China}

\author[0000-0002-9615-1481]{Hui Sun}
\affiliation{Key Laboratory of Space Astronomy and Technology, National Astronomical Observatories, Chinese Academy of Sciences, Beijing 100012, People's Republic of China}

\author[0000-0001-6374-8313]{Yuan-Pei Yang}
\affiliation{South-Western Institute for Astronomy Research, Yunnan University, Kunming, Yunnan 650500, People’s Republic of China; \url{ypyang@ynu.edu.cn}}

\author{Zhuo Li}
\affiliation{Department of Astronomy, School of Physics, Peking University, Beijing 100871, China; \url{zhujp@pku.edu.cn}}
\affiliation{Kavli Institute for Astronomy and Astrophysics, Peking University, Beijing 100871, China; \url{zhuo.li@pku.edu.cn}}

\author[0000-0002-6442-7850]{Rui-Chong Hu}
\affiliation{Guangxi Key Laboratory for Relativistic Astrophysics, School of Physical Science and Technology, Guangxi University, Nanning 530004, China}

\author[0000-0002-2956-8367]{Ying Qin}
\affiliation{Department of Physics, Anhui Normal University, Wuhu, Anhui, 241000, China}

\author[0000-0002-9188-5435]{Shichao Wu}
\affiliation{Max-Planck-Institut f{\"u}r Gravitationsphysik (Albert-Einstein-Institut), D-30167 Hannover, Germany}
\affiliation{Leibniz Universit{\"a}t Hannover, D-30167 Hannover, Germany}

\begin{abstract}

Gamma-ray bursts (GRBs) have been phenomenologically divided into long- and short-duration populations, generally corresponding to the collapsar and compact merger origin, respectively. Here we collect three unique bursts, GRBs\,060614, 211211A and 211227A, all characterized by a long-duration main emission (ME) phase and a rebrightening extended emission (EE) phase, to study their observed properties and the potential origin as neutron star--black hole (NSBH) mergers. NS-first-born (BH-first-born) NSBH mergers tend to contain fast-spinning (non-spinning) BHs that more easily (hardly) allow tidal disruption to happen with (without) forming electromagnetic signals. We find that NS-first-born NSBH mergers can well interpret the origins of these three GRBs, supported by that: (1) Their X-ray MEs and EEs show unambiguous fall-back accretion signatures, decreasing as $\propto{t}^{-5/3}$, which might account for their long duration. The EEs can result from the fall-back accretion of $r$-process heating materials, predicted to occur after NSBH mergers. (2) The beaming-corrected local event rate density for this type of merger-origin long-duration GRBs is $\mathcal{R}_0\sim2.4^{+2.3}_{-1.3}\,{\rm{Gpc}}^{-3}\,{\rm{yr}}^{-1}$, consistent with that of NS-first-born NSBH mergers. (3) Our detailed analysis on the EE, afterglow and kilonova of the recently high-impact event GRB\,211211A reveals it could be a merger between a $\sim1.23^{+0.06}_{-0.07}\,M_\odot$ NS and a $\sim8.21^{+0.77}_{-0.75}\,M_\odot$ BH with an aligned-spin of $\chi_{\rm{BH}}\sim0.62^{+0.06}_{-0.07}$, supporting an NS-first-born NSBH formation channel. Long-duration burst with rebrightening fall-back accretion signature after ME, and bright kilonova might be commonly observed features for on-axis NSBH mergers. We estimate the multimessenger detection rate between gravitational waves, GRB and kilonova emissions from NSBH mergers in O4 (O5) is $\sim0.1\,{\rm{yr}}^{-1}$ ($\sim1\,{\rm{yr}}^{-1}$).

\end{abstract}


\section{Introduction} \label{sec:intro}

In observations, it is usually adopted that a critical duration of $T_{90}\sim2\,{\rm{s}}$ separates gamma-ray bursts (GRBs) into long- and short-duration populations \citep{norris1984,kouveliotou1993}. Long-duration GRBs (lGRBs) have been identified to be originated from massive collapsar by their association with broad-line Type Ic supernovae \citep[e.g.,][]{galama1998,woosley2006} and their exclusive hosts in star-forming galaxies \citep[e.g.,][]{bloom1998,christensen2004}. It has long been suspected that neutron star mergers, including binary neutron star (BNS) and neutron star--black hole (NSBH) mergers, are potential origins of short-duration GRBs \citep[sGRBs;][]{paczynski1986,paczynski1991,eichler1989,narayan1992}. Due to natal kicks impacted to the binaries at birth and long inspiral delays before mergers, NS mergers are believed to occur in low-density environments with significant offsets away from the centers of their host galaxies \citep[e.g.,][]{narayan1992,bloom1999} supported by observations \citep[e.g.,][]{fong2010,fong2015,li2016}. NS mergers can release an amount of neutron-rich matter \citep{lattimer1974,lattimer1976,sybalisty1982} that allows elements heavier than iron to be synthesized via the rapid neutron-capture process ({\em{r}}-process). It was predicted that the radioactive decay of these {\em{r}}-process nuclei would power an ultraviolet-optical–infrared thermal transient named ``kilonova'' \citep{li1998,metzger2010}.

The smoking-gun evidence for the BNS merger origin of sGRB and kilonova was the multimessenger observations of the first BNS merger gravitational-wave (GW) source GW170817 detected by the LIGO/Virgo Collaboration \citep[LVC;][]{abbott2017gw170817} and subsequent associated electromagnetic (EM) signals, including an sGRB GRB\,170817A triggered by the Fermi Gamma-ray Burst Monitor \citep{abbott2017gravitational,goldstein2017,savchenko2017,zhangbb2018}, a broadband jet afterglow from radio to X-ray with an off-axis viewing angle \citep[e.g.,][]{margutti2017,troja2017,lazzati2018,lyman2018,lamb2019,ghirlanda2019} and a fast-evolving kilonova transient \citep[AT2017gfo; e.g.,][]{abbott2017multimessenger,arcavi2017,coulter2017,drout2017,evans2017,kasliwal2017,kilpatrick2017,pian2017,smartt2017}. With the confirmation of the origin of sGRB and kilonova from the BNS merger population, one may especially expect to further establish the connection between NSBH mergers and their associated EM counterparts. However, although {{two high-confidence NSBHs (i.e., GW200105 and GW200115) and}} a few {{marginal}} NSBH GW candidates were detected during the third observing run of LVC \citep{abbott2021nsbh,nitz2021,abbott2021population}, EM counterparts by the follow-up observations of these GWs were missing \citep[e.g.,][]{anand2021,andreoni2020,coughlin2020,gompertz2020,kasliwal2020,page2020}, except an amphibious association between a subthreshold GRB GBM-190816 and a subthreshold NSBH event \citep{goldstein2019,yang2020}. One plausible explanation for the lack of detection of an EM counterpart is that present EM searches were too shallow to achieve distance and volumetric coverage for the probability maps of LVC events \citep{saguescarracedo2021,coughlin2020,zhu2021kilonova}. Furthermore, detailed studies on these NSBH candidates \citep{zhu2021no,zhu2022,fragione2021,mandel2021,gompertz2022,dorazio2022} revealed that they were more likely to be plunging events and could hardly produce any bright EM signals owing to near-zero spins of the primary BHs, since NSBH mergers tend to make tidal disruptions and drive bright EM counterparts if the primary BHs have high aligned-spins \citep[e.g.,][]{kyutoku2015,foucart2018,zhu2021no,zhu2022,diclemente2022}.

Due to the lack of smoking-gun evidence, it is unclear whether NSBH mergers can contribute to the sGRB population \citep[e.g.,][]{gompertz2020searching}. On the one hand, the majority of NSBH binaries are believed to originate from the {{classic isolated binary evolution scenario (involving a common-envelope)}} \citep[e.g.,][]{giacobbo2018,belczynski2020,drozda2020,shao2021}. In this scenario, the primary BHs are usually born first and have negligible spins consistent with the properties of LVC NSBH candidates \citep{broekgaarden2021,zhu2021no}. Conversely, if the NSs are born first, the progenitors of the BHs would be tidally spun up efficiently by the NSs {{in close binaries (orbital periods $\lesssim2\,{\rm d}$)}} and finally form fast-spinning BHs \citep{hu2022}. {{A fractional of these NS-first-born NSBH systems formed in close binaries can merge within Hubble time.}} Therefore, compared with BH-first-born NSBH mergers, NS-first-born NSBH mergers are easier to allow tidal disruption to happen and drive bright GRB emissions. Because NS-first-born NSBH mergers may only account for $\lesssim20\%$ NSBH populations \citep{romangarza2021,chattopadhyay2021,chattopadhyay2022}, GRB populations contributed from NSBH mergers should be limited. On the other hand, most disrupted NSBH mergers can eject much more materials and lead to more powerful fall-back accretions than BNS mergers \citep{rosswog2007,fernandez2017}. Furthermore, $r$-process heating might affect the fall-back accretion of marginally bound matter \citep{metzger2010fb}. A late-time fall-back accretion of these materials may happen after tens of seconds of the merger if the remnant BH has a mass of $\gtrsim6-8\,M_\odot$. Because most NSBH mergers can remain BHs with masses in this range, \cite{desai2019} suggested that an extended emission (EE) caused by the fall-back accretion of $r$-process heating materials can be an important signal to distinguish NSBH GRBs from BNS GRBs. Thus, it is plausible that the energy budgets, durations, and other observed properties of NSBH GRBs could differ from those of BNS mergers.

Very recently, the observations of an lGRB (i.e., GRB\,211211A) associated with a kilonova emission at a redshift $z = 0.0763$ (luminosity distance $D_{\rm L} \approx 350\,{\rm Mpc}$) was reported by a few groups \citep{rastinejad2022,yang2022,xiao2022,gompertz2022grb,mei2022,zhang2022,chang2022}. The burst was characterized by a spiky main emission (ME) phase with a duration of $\sim13\,{\rm{s}}$, an EE phase lasting $\sim55\,{\rm{s}}$, and a temporal lull between these two phases. Since the observation property of its associated kilonova emission was similar to that of AT2017gfo\footnote{\cite{waxman2022} suggested that the burst could happen in another spatially nearby galaxy at a higher redshift. The near infrared emission following GRB\,211211A could be thermal emission from dust, heated by UV radiation produced by the interaction between the jet plasma and the cirumstellar medium, rather than a kilonova emission.} \citep{rastinejad2022,xiao2022}, indicating an origin of a compact binary coalescence, it was a challenge to interpret the intrinsically long duration of the burst. \cite{yang2022} proposed that a merger of a near-equal-mass NS--white dwarf binary can well explain the ME of GRB\,211211A, since the accretion of some high-angular-momentum white dwarf debris onto the remnant NS can prolong the burst duration. \cite{gao2022} suggested a strong magnetic flux may surround the central engine of GRB\,211211A, which results in the long-time accretion process due to the magnetic barrier effect \citep{proga2006,liu2012}.

Besides GRB\,211211A, two other redshift-known ($z$-known) lGRBs, i.e., GRB\,060614 and GRB\,211227A, were proposed to derive from compact binary coalescences. GRB\,060614 \citep{galyam2006,dellavalle2006,zhang2007} were found to be associated with a kilonova candidate \citep{yang2015}, while GRB\,211227A showed a large physical offset from the host center and lacked a supernova signature that should have been observed at the location of the burst \citep{Lv2022}. In this {\em Letter}, we study the properties of these three merger-origin lGRBs, especially for GRB\,211211A, and show that a single explosive population via the NS-first-born NSBH merger can account for their origin. Here, the cosmological parameters are taken as $H_0 = 67.4\,{\rm km}\,{\rm s}^{-1}\,{\rm Mpc}^{-1}$, $\Omega_{\rm m} = 0.315$, $\Omega_\Lambda = 0.685$ \citep{planck2020}.

\section{Properties of Merger-origin LGRBs}

\begin{table*}[]
\begin{center}
\begin{threeparttable}
\caption{The observed properties of GRB\,211211A, GRB\,060614\tnote{1}, and GRB\,211227A\tnote{2} \label{tab:1}}
\begin{tabular}{ccccc}
\hline
\hline
& GRB\,211211A & GRB\,060614 & GRB\,211227A \\
\hline
\textbf{Main Emission} & & & \\
Duration (s) & 13 & 6 & 4 \\
Peak energy (keV) & $687_{-11}^{+13}$ & $300_{-90}^{+210}$ & $400_{-200}^{+1200}$ \\
Energy fluence ($\rm erg\,cm^{-2}$) & $3.77_{-0.01}^{+0.01}\times10^{-4}$ & $8.2_{-2.5}^{+0.6}\times10^{-6}$ & $2.01_{-0.42}^{+0.19}\times10^{-6}$ \\
Isotropic equivalent energy (erg) & $5.30^{+0.01}_{-0.01}\times10^{51}$ & $3.18_{-0.98}^{+0.22}\times10^{50}$ & $2.69_{-0.56}^{+0.25}\times10^{50}$ \\
Spectral index $\alpha$ & $-0.996_{-0.005}^{+0.005}$ & $-1.57_{-0.14}^{+0.12}$ & $-1.56_{-0.06}^{+0.15}$\\
Spectral index $\beta$ & $-2.36_{-0.02}^{+0.02}$ & ... & ... \\
\hline
\textbf{Whole Emission} & & & \\
Duration (s) & $43.18_{-0.06}^{+0.06}$ & $102\pm5$ & 84 \\
Peak energy (keV) & $399_{-16}^{+14}$ & $10-100$ & $192_{-42}^{+45}$ \\
Energy fluence\tnote{3} ($\rm erg\,cm^{-2}$) & $5.42_{-0.08}^{+0.08}\times10^{-4}$ & $4.09_{-0.34}^{+0.18}\times10^{-5}$ & $2.60_{-0.21}^{+0.21}\times10^{-5}$ \\
Isotropic equivalent energy (erg) & $7.61^{+0.11}_{-0.11}\times10^{51}$ & $1.59_{-0.13}^{+0.07}\times10^{51}$ & $3.48_{-0.16}^{+0.16}\times10^{51}$ \\
Spectral index $\alpha$ & $-1.20_{-0.01}^{+0.01}$ & ... & $-1.34_{-0.08}^{+0.10}$ \\
Spectral index $\beta$ & $-2.05_{-0.02}^{+0.02}$ & ... & $-2.26_{-1.11}^{+0.24}$ \\
\hline
Redshift & 0.076 & 0.125 & 0.228 \\
\hline
\end{tabular}
\begin{tablenotes}
\item[1]The data of GRB\,211221A and GRB\,060614 are collected from Table 1 of \cite{yang2022}.
\item[2]The data of GRB\,211227A are collected from \cite{Lv2022} and \cite{2022GCN.31544....1T}.
\item[3]The energy fluence of the ME for GRB\,211227A is calculated by the \texttt{HEASoft} tool \citep{heasoft2014} in the $15-1500\,{\rm keV}$ energy band.
\end{tablenotes}
\end{threeparttable}
\end{center}
\end{table*}

\subsection{Observed Properties and X-ray Fall-back Accretion Signals}

\begin{figure*}
    \centering
	\includegraphics[width = 0.49\linewidth , trim = 70 30 20 30, clip]{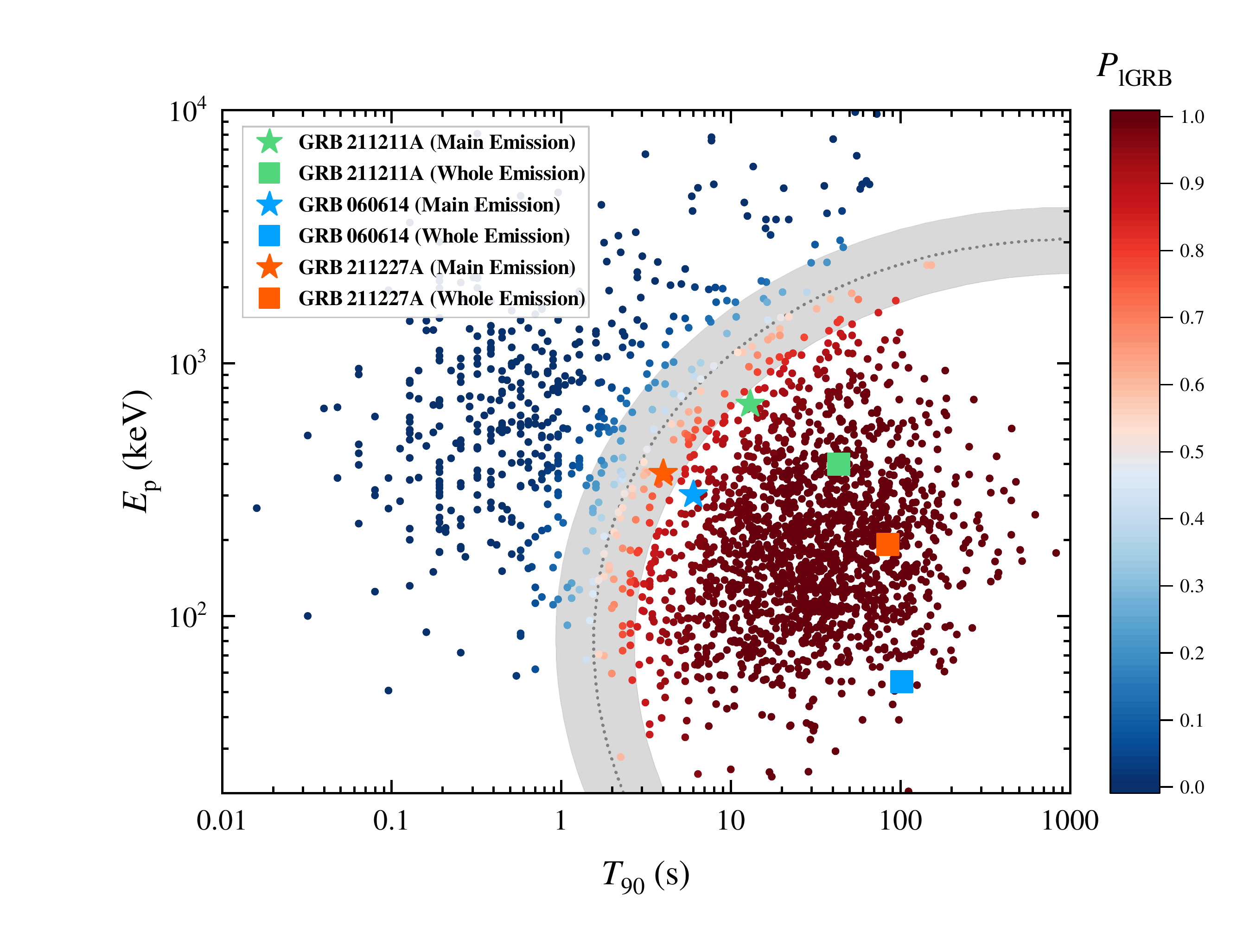}
	\includegraphics[width = 0.49\linewidth , trim = 70 30 20 30, clip]{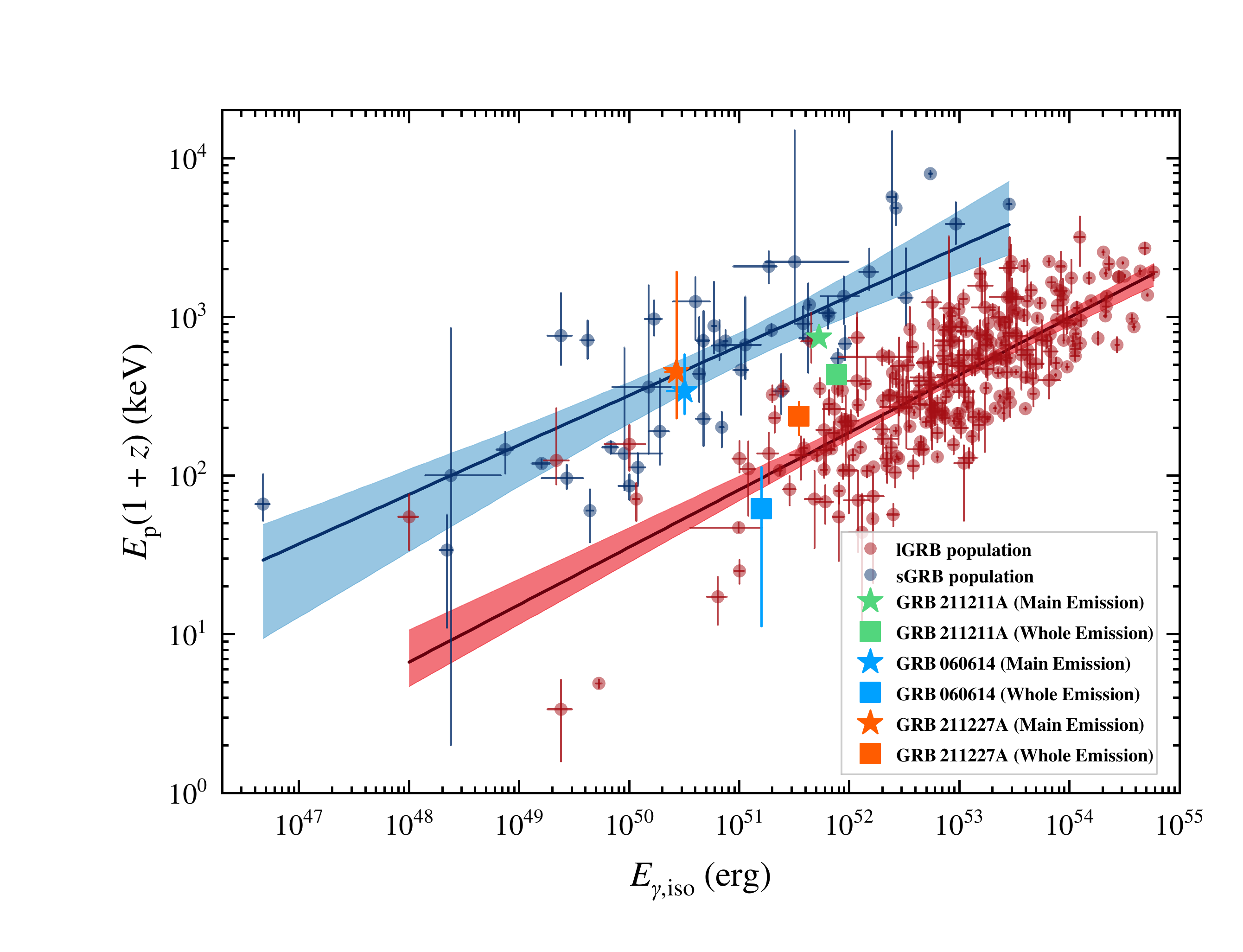}
    \caption{{Left Panel:} Long/short classification diagram in the $T_{\rm 90}$ - $E_{\rm p}$ domain \citep{Kienlin2014, Gruber2014, Bhat2016, Kienlin2020}. The dashed line and gray shaded region are the best-fit and 1$\sigma$ credible boundaries to extinguish lGRBs from sGRBs, respectively. The redder (bluer) the color of the point, the higher the possibility of the lGRB (sGRB) origin. {Right Panel:} $E_{\rm p}(1+z)$ and $E_{\rm \gamma, iso}$ correlation diagram with known redshift data \citep{ amati2002, zhangbing2009, minaev2020}. The red and blue solid lines represent the best-fit correlations for lGRB and sGRBs, respectively. The green, blue, and orange stars (squares) in both panels represent the placement of ME (WE) for GRB\,211211A, GRB\,060614, and GRB\,211227A, respectively. }
    \label{fig:Properties}
\end{figure*}

We collect the observed data of three present $z$-known merger-origin lGRBs, including GRB\,211221A \citep[e.g.,][]{rastinejad2022,yang2022,xiao2022}, GRB\,060614 \citep{galyam2006,dellavalle2006,zhang2007}, and GRB\,211227A \citep{Lv2022}, all characterized by a spiky long-duration ME phase, a rebrightening EE phase, and a temporal lull between these two phases, to study their similarities. {{Here, the EE phase is defined as the long-lasting lower level emission phase after the initial intense ME phase \citep{Norris2006,Lanlin2020}. The whole emission (WE) phase includes the ME phase and the EE phase. }}  Table \ref{tab:1} lists the observed properties of the MEs and the WEs.

In the left panel of Figure \ref{fig:Properties}, we divide detected GRBs into two populations, i.e., lGRB and sGRB populations, in the duration $T_{90}$ vs. Earth-frame peak energy $E_{\rm p}$ diagram through the Gaussian mixture model. {{The intermediate population at the boundary region between lGRB and sGRB populations could be originated from collaspar or compact object coalescence \citep[e.g.,][]{tunicliffe2012,zaninoni2016}.}}  Both MEs and WEs of these three merger-origin GRBs fall into the distribution of lGRBs. Therefore, without the redshift, host galaxy information, and associated kilonova detection, these three bursts would be classified as members of the lGRB population due to their long durations.

The correlations between the total isotropic equivalent energy $E_{\gamma,\rm iso}$ of the prompt emission and the rest-frame peak energy $E_{\rm p}(1+z)$ \citep[Amati relation;][]{amati2002} for both lGRB and sGRB populations are shown in the right panel of Figure \ref{fig:Properties}. We find that the WEs of these three bursts still behave as normal lGRBs, although their $E_{\gamma,\rm{iso}}$ are lower than most observed lGRBs. Conversely, their MEs lie on the middle location of the sGRB track rather than on the lGRB track. The difference of $E_{\gamma,\rm iso}$ between WEs and MEs, corresponding to the $E_{\gamma,\rm iso}$ of the EEs, for these bursts are similar, which are $\sim 1-2\times10^{51}\,{\rm erg}$.

\begin{figure}
    \centering
	\includegraphics[width = 1\linewidth , trim = 70 30 90 40, clip]{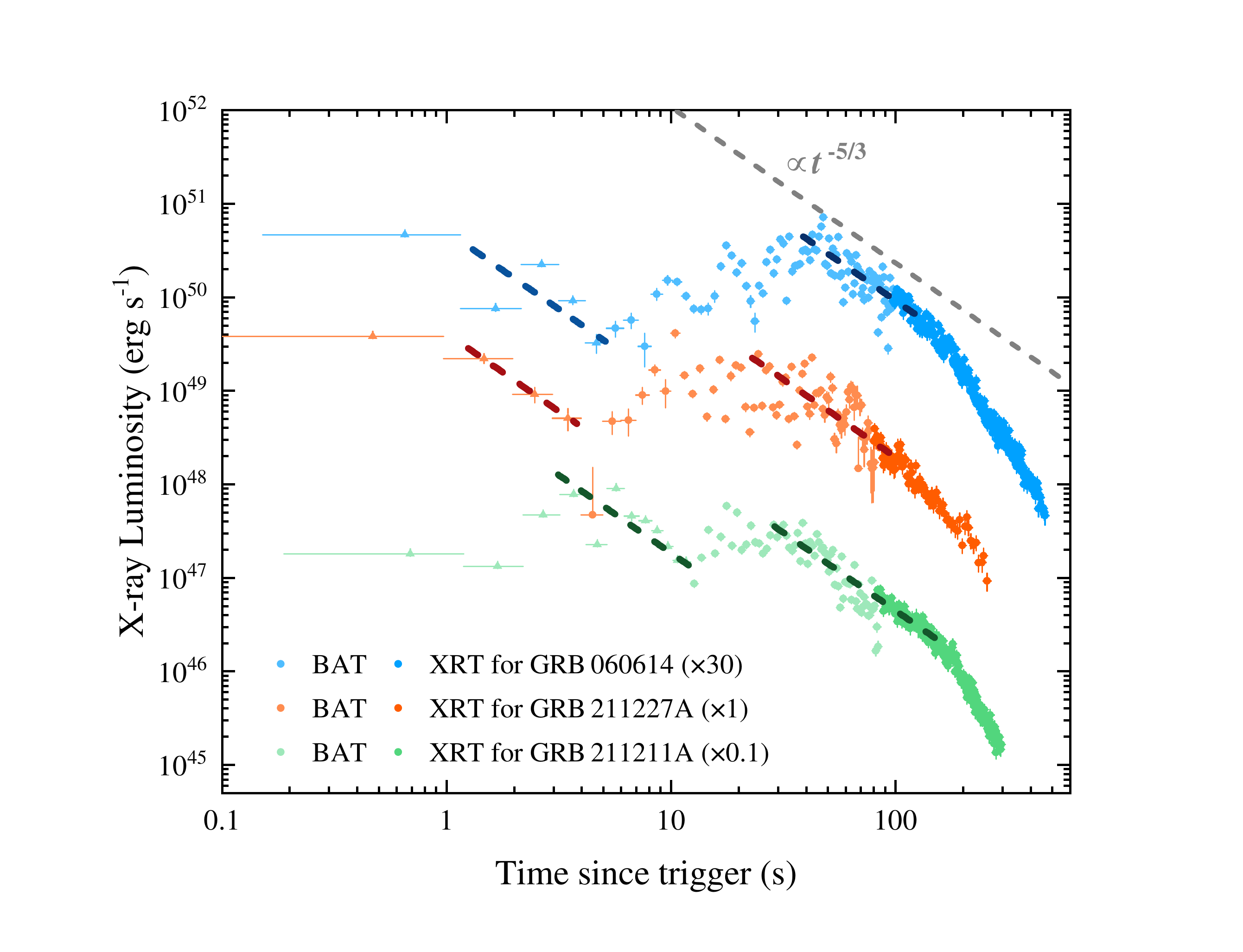}
    \caption{BAT and XRT light curves of GRB\,211211A (green), GRB\,060614 (blue) and GRB\,211227A (orange). BAT luminosity is calculated at 10\,keV. {{Their MEs and EEs are marked as triangle and circle points, respectively.}} The dashed lines represent the X-ray light curves track the $\propto t^{-5/3}$ mass fall-back accretion.}
    \label{fig:Xray}
\end{figure}

After a BNS/disrupted NSBH merger, a compact remnant accretion disk would be formed around the NS or BH. The lifetime of this disk is typically $\lesssim1\,{\rm s}$, which is thought to determine the burst duration of an sGRB \citep[e.g.,][]{shapiro2017,zhang2019,ruiz2020}. In order to explain the long durations of these three merger-origin GRBs, an additional energy/matter injection is needed. When a group of bound ejecta with an energy distribution of $dM/dE \propto E^\alpha$ where $\alpha \approx 0$ around $E = 0$ fall back onto the central NS or BH, the fall-back rate would track as $\propto t^{-5/3}$ \citep{rees1988}. As shown in Figure \ref{fig:Xray}, we find that there are long-duration fall-back accretion signals {{appearing}} in the X-ray light curves of both MEs and EEs for these three bursts. The fall-back rates in their EEs peak at $\sim20-40\,{\rm s}$ and follow $\propto t^{-5/3}$ with a duration of $\sim100-200\,{\rm s}$. Thus, the fall-back accretion might result in their long duration.

\begin{figure}
    \centering
	\includegraphics[width = 1\linewidth , trim = 70 30 90 40, clip]{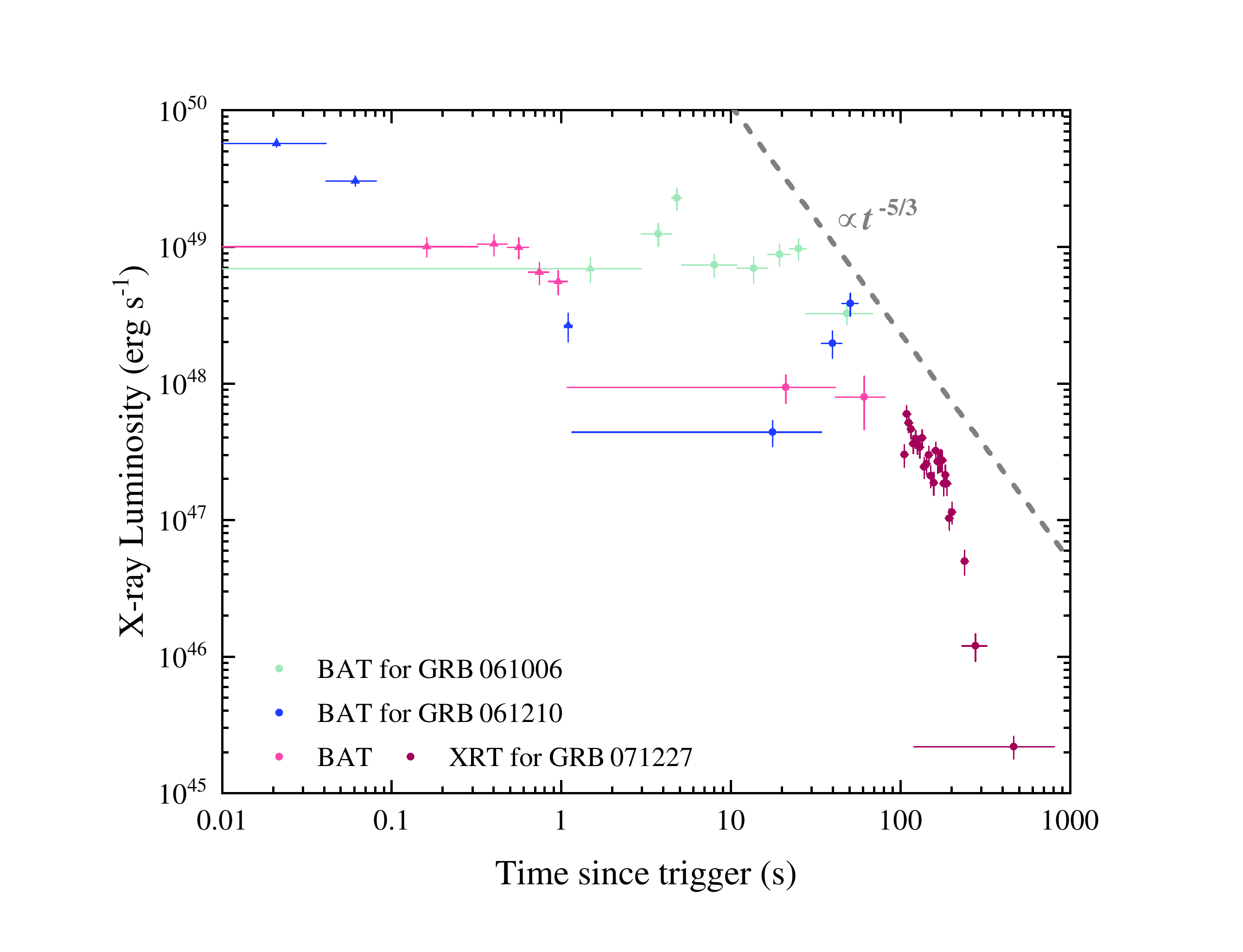}
    \caption{Similar to Figure \ref{fig:Xray}, but for GRB\,061006 (green), GRB\,061210 (dark blue) and GRB\,071227 (purple).}
    \label{fig:Xray2}
\end{figure}

\cite{gompertz2020} collected 39 $z$-known sGRBs (7 events had {{an}} EE) and searched for possible NSBH candidates among these sGRBs. However, they didn't find any clear evidences for the existence of NSBH GRBs in this complete sGRB sample. {{If a millisecond magnetar can survive after a BNS merger, the magnetar can {{lose}} its rotational energy via the spin-down process and result in a long-duration X-ray plateau in the X-ray lightcurve \citep{zhang2001}. \cite{gompertz2020} presented that 4 of these sGRBs with EEs in their samples could be magnetar-like, so they cannot be originated from NSBH mergers.}} In Figure \ref{fig:Xray2}, we collect the X-ray light curve of other 3 sGRBs with EEs, including GRB\,061006 \citep{berger2007}, GRB\,061210 \citep{berger2007} and GRB\,071227 \citep{davanzo2009}, whose X-ray emissions could not be well interpreted via the magnetar model by \cite{gompertz2020}. We find that GRB\,071227 could still be magnetar-like, characterized by a long-duration X-ray plateau. {{The X-ray lightcurves of GRB\,061006 and GRB\,061210 do not show any signatures of $\propto t^{-5/3}$ decay and hence may provide no evidence of unambiguous fall-back accretion signal, though that could be partly because of their limited data points}}. Thus, the EEs in present $z$-known sGRBs may not be caused by the fall-back accretions. The similar observed properties of merger-origin lGRBs, {{i.e., long durations and unambiguous rebrightening fall-back accretion signals, do not present in the existent observations of}} those of merger-origin sGRBs with/without EEs, indicating an unique origin for them.

In principle, both BNS and NSBH mergers can generate the early-time fall-back accretions \citep{rosswog2007,metzger2010fb,fernandez2017}. However, most of disrupted NSBH mergers would eject much more materials and lead to more powerful fall-back accretions than BNS mergers \citep{rosswog2007}. Furthermore, the simulations by \cite{metzger2010fb} and \cite{desai2019} showed that $r$-process heating might affect on the materials that is marginally gravitationally bound. {{The gravity of the remnants formed after BNS mergers might be too low to drag these $r$-process heating materials back and result in fall-back accretions.}} However, as predicted by \cite{desai2019}, a late-time fall-back accretion of these materials may happen after tens of seconds of the merger if the remnant BH has a mass of $\gtrsim6-8\,M_\odot${{, leading to a rebrightening emission with $t^{-5/3}$ power-law decay appeared in the X-ray lightcurve.}} The EEs of these three bursts, {{whose starting times and observational features are consistent with the predictions of \cite{desai2019},}} may be originated from the fall-back accretion of $r$-process heating materials. We thus suspect that these three bursts are derived from NSBH mergers.

\subsection{Event Rate Density} \label{sec:EventRate}

Assuming the unique merger origin of the three lGRBs, we investigate their local event rate density by tentatively constructing the luminosity function (LF) following \cite{sun2015,sun2022}. The limited number of events can lead to large uncertainty of the LF. However, it can be plausible to make a fine evaluation by adding up the contribution of all three events. Following the Equation (12) in \cite{sun2022}, the LF $\Phi(L)$ is the sum of each event $j$ in the range from $\log_{10} L$ to $\log_{10} L + d\log_{10} L$  (with total $\Delta N_{L}$ events). 
\begin{equation}
\begin{split}
\Phi(L)d \log_{10} L &= \sum_{j=1}^{\Delta N_{L}} \left[ \Phi(L)d \log_{10} L \right]_j \\
& = \sum_{j=1}^{\Delta N_{L}} \frac{1}{\left[ T_{\rm BAT} \times V'_{\rm max}\right]_j}, 
\end{split}
\label{eq:lf}
\end{equation}
where $T_{\rm BAT}$ is the total monitoring time by Swift/BAT and $V'_{\rm max}$ is the effective maximum volume that is monitored  (with field of view $\Omega_{\rm BAT}$) weighted by the density evolution $f(z)$ and time dilation, i.e.,
\begin{equation}
V'_{\rm max} = \int_{0}^{z_{\rm max}} \frac{\Omega_{\rm BAT}}{4\pi} \cdot \frac{f(z)}{1+z} \frac{dV(z)}{dz} dz.
\label{eq:V'max}
\end{equation}
The $z_{\rm max}$ is the maximum redshift that an event can be detected with bolometric peak luminosity $L_{\rm bol}$. The bolometric peak luminosity of these three events, defined in the energy range of $1 - 10^4\,\rm keV$, are derived from $k$-correction based on the spectrum listed in Table \ref{tab:1} following Equation (29) in \cite{sun2015}.

We take the instrument parameters of Swift/BAT with $\Omega_{\rm BAT} = 1.3$, $T_{\rm BAT} = 17\,\rm yr$ and the flux sensitivity $f_{\rm th, BAT} = 3 \times 10^{-8}\,\rm erg\,cm^{-2}\,s^{-1}$. The $f(z)$ for GRBs with merger-origin is adopted from \cite{zhu2021kilonova}. We take the correction of the redshift measurement into account. For both the lGRBs and sGRBs sample, the ratio between the total number and the number of $z$-known events from the Swift observations is approximately $4:1$. In addition, since Swift has relatively softer energy band than BATSE, it tends to detect more lGRBs than sGRBs. One needs to adopt another correction by a factor of 3 to compensate the short (hard)-to-long (soft) ratio of Swift-detected GRBs in comparison to those by BATSE observations \citep{sun2015}. 

We divided the events into two bins with $\Delta \log_{10} L = 1$ and fitted the LF with a single power law of the slope $0.6 \pm 0.8$. The local event rate density, which is derived by integrating the LF above the luminosity threshold $3 \times 10^{50}\,\rm erg\,s^{-1}$, is given as 
\begin{equation}
{\mathcal R}_{0} (> 3 \times 10^{50}\,\rm erg\,s^{-1}) = 2.4 ^{+2.3}_{-1.3} \times 10^{-2} \, \rm Gpc^{-3}\,yr^{-1}.
\end{equation}
The errors are given at the $1\sigma$ confidence level \citep{gehrels1986}.

The intrinsic local event rate densities of both lGRBs and sGRBs are of the order of unity in unit of $\rm Gpc^{-3}\,yr^{-1}$ above the isotropically bolometric luminosity of $\sim 10^{50}\,\rm erg\,s^{-1}$ \citep{sun2015}. We find that the local event rate density for the lGRBs with merger origin is much lower than that of both lGRBs or sGRBs. The rate density for BNS inferred using GWs through GWTC-2 (GWTC-3) is ${320}_{-240}^{+490}\,{\rm{Gpc}}^{-3}\,{\rm{yr}}^{-1}$ ($10-1700\,{\rm Gpc}^{-3}\,{\rm yr}^{-1}$) \citep[e.g.,][]{abbott2021population1,abbott2021population,mandel2022}. {{\cite{abbott2021nsbh} inferred a NSBH rate density of ${45}_{-33}^{+75}\,{\rm{Gpc}}^{-3}\,{\rm{yr}}^{-1}$ by considering the observations of two NSBH mergers or $130^{+112}_{-69}\,{\rm{Gpc}}^{-3}\,{\rm{yr}}^{-1}$ assuming a broad NSBH population, while GWTC-3 \citep{abbott2021population} reported the NSBH merger rate density to be between $7.8-140\,{\rm Gpc}^{-3}\,{\rm yr}^{-1}$.}} By considering a jet beaming factor of $f_{\rm b} = 0.01$, the beaming-corrected event rate density for merger-origin lGRBs is ${\mathcal R}_{0,\rm b}=2.4^{+2.3}_{-1.3}(f_{\rm b}/0.01)\,\rm Gpc^{-3}\,yr^{-1}$, which is much lower than those of BNS and NSBH mergers. {{However,}} the beaming-corrected rate density of merger-origin lGRBs is consistent with that of NS-first-born NSBH merger \citep[$\lesssim20\%$ of total NSBH populations;][]{romangarza2021,chattopadhyay2022}. {{By investigating the parameter space of forming NS-first-born NSBH binaries, \cite{hu2022} found that most of NSBH binaries that can merge within Hubble time would have BHs with projected aligned spins $\chi_{\rm BH}\gtrsim0.8$ and, hence, can certainly make tidal disruptions to produce electromagnetic counterparts. Only a small fractional low-mass BHs with $\chi_{\rm BH}\sim0.2-0.8$ can merge with an NS within Hubble time and can still allow tidal disruption to happen if NSs are not really massive (i.e., $M_{\rm NS}\gtrsim1.6-2.0\,M_\odot$).}} Since the rest of BH-first-born NSBH mergers would mostly contribute to plunging events due to negligible projected aligned-spins of BH components \citep{zhu2022}, a single explosive population via the NS-first-born NSBH merger can account for their origin.

\section{Modeling and Origin of GRB\,211211A}

{{GRB211211A, as a recently high-impact event, had one of the most complete multi-band data records of afterglow and kilonova, which can give a strict constraint on our fitting parameters to explore its plausible origin.}} In this section, we will simultaneously interpret the emissions of $\gamma$-ray/X-ray EEs by the fall-back accretion of $r$-process heating materials, afterglow emissions, and kilonova emissions of GRB\,211211A, within the framework of NSBH mergers. Since the structures of the prompt emissions of GRBs are generally believed to originate from the internal shock processes, in the following discussion, we are only interested in the light curve outline that mainly depends on the engine power due to the fall-back accretion.

\subsection{Modeling}

\subsubsection{Fall-back Accretion of $r$-process Heating Materials}

For NSBH mergers with NS tidal disruption, a fractional of $r$-process heating materials would fall back onto the remnant BH tens of seconds after the merger, resulting in the EE \citep{metzger2010fb,desai2019} through the Blandford-Znajek (BZ) mechanism \citep{blandford1977}. Following \cite{macfadyen2001} and \cite{dai2012}, the fall-back rate initially increases with time as $\dot{M}\propto t^{1/2}$ before the time of $t_{\rm p}$ corresponding to the peak fall-back rate $\dot{M}_{\rm p}$. Then, the late-time fall-back accretion behavior would track as $\dot{M}\propto t^{-5/3}$ until the break time $t_{\rm b}$ \citep{chevalier1998}. While most of fall-back materials are accreted after $t_{\rm b}$, we describe the fall-back rate as $\dot{M}\propto t^{-s}$. An empirical three-segment broken power-law function is adopted to model the fall-back accretion rate of $r$-process heating materials, i.e.,
\begin{equation}
\label{equ:fb}
\begin{split}
    \dot{M}(t) =  &\dot{M}_{\rm p}\left[\frac{1}{2}\left(\frac{t-t_0}{t_{\rm p}-t_0}\right)^{-1/2} + \frac{1}{2}\left(\frac{t-t_0}{t_{\rm p}-t_0}\right)^{5/3}\right]^{-1}\\
    &\times\left[1 + \frac{1}{2}\left(\frac{t-t_0}{t_{\rm b}-t_0}\right)^{s-5/3}\right]^{-1},
\end{split}
\end{equation}
where $t_0$ is the starting time of the fall-back accretion.

The BZ power is related to the mass and spin of the central BH \citep[e.g.,][]{li2000,wu2013,lei2013,lei2017,liu2017}, i.e.,
\begin{equation}
    L_{\rm BZ} = 1.7\times10^{50}\,{\rm erg}\,{\rm s}^{-1}\chi_{\rm BH}\left(\frac{M_{\rm BH}}{M_\odot}\right)^2B_{\rm BH,15}^2F(\chi_{\rm BH}),
\end{equation}
where $M_{\rm BH}$ is the central BH mass, $\chi_{\rm BH}$ is the dimensionless aligned-spin of the BH, $B_{\rm BH,15} = B_{\rm BH}/10^{15}\,{\rm G}$ is the magnetic field strength threading the BH horizon and $F(\chi_{\rm BH}) = [(1+q^2)/q^2][(q+1/q)\arctan q-1]$ with $q = \chi_{\rm BH}/(1+\sqrt{1-\chi_{\rm BH}^2})$. The magnetic field can be estimated through the balance between the magnetic pressure on the horizon and the ram pressure of the innermost part of the accretion flow:
\begin{equation}
    \frac{B_{\rm BH}^2}{8\pi} = P_{\rm ram} \sim \rho c^2\sim\frac{\dot{M}c}{4\pi r_{\rm H}^2},
\end{equation}
where $c$ is the speed of light and $r_{\rm H} = (1+\sqrt{1-\chi_{\rm BH}^2})r_{\rm g}$ is the radius of the BH horizon with the Schwarzschild radius $r_{\rm g} = GM_{\rm BH}/c^2$ and the gravitational constant $G$. Thus, the BZ power can be also expressed as
\begin{equation}
\label{equ:BZeffect}
    L_{\rm BZ} = 9.3\times10^{53}\,{\rm erg}\,{\rm s}^{-1}\frac{\chi_{\rm BH}^2F(\chi_{\rm BH})}{(1 + \sqrt{1-\chi_{\rm BH}^2})^2}\frac{\dot{M}}{M_\odot\,{\rm s}^{-1}}.
\end{equation}

Because the central BH would be spun up by accretion and spun down by the BZ mechanism, the conservation of energy and angular momentum of a BH can be written as
\begin{equation}
\begin{split}
\label{equ:dM_BHdt}
    \frac{dM_{\rm BH}c^2}{dt} &= \dot{M}c^2E_{\rm ISCO} - L_{\rm BZ}, \\
    \frac{dJ_{\rm BH}}{dt} &= \dot{M}J_{\rm ISCO} - 2L_{\rm BZ}/\Omega_{\rm H},
\end{split}
\end{equation}
where $\Omega_{\rm H}=c\chi_{\rm BH}/(2r_{\rm H})$ is the angular velocity of the BH horizon, $E_{\rm ISCO} = (4\sqrt{\widetilde{R}_{\rm ISCO}}-3\chi_{\rm BH})/\sqrt{3}\widetilde{R}_{\rm ISCO}$ and $J_{\rm ISCO} = 2GM_{\rm BH}(3\sqrt{\widetilde{R}_{\rm ISCO}} - 2\chi_{\rm BH})/c\sqrt{3\widetilde{R}_{\rm ISCO}}$ are the specific energy and specific angular momentum of a particle at the innermost stable circular orbit (ISCO) radius \citep{novikov1973}, respectively. $\widetilde{R}_{\rm ISCO} = 3 + Z_2 - {\rm {sign}}(\chi_{\rm {BH}})\sqrt{(3 - Z_1)(3 + Z_1 + 2Z_2)}$ represents the normalized radius of the BH ISCO with $Z_1 = 1 + (1 - \chi_{\rm {BH}}^2) ^ {1 / 3} [(1 + \chi_{\rm {BH}})^{1 / 3} + (1 - \chi_{\rm {BH}})^{1 / 3}]$ and $Z_2 = \sqrt{3 \chi_{\rm {BH}}^2 + Z_1^2}$ \citep{bardeen1972}. Since the angular momentum of BH is expressed as $J_{\rm BH} = GM_{\rm BH}^2\chi_{\rm BH}/c$, one has
\begin{equation}
\begin{split}
\label{equ:dchi_BHdt}
    \frac{d\chi_{\rm BH}}{dt} =& \frac{(\dot{M}J_{\rm ISCO} - 2L_{\rm BZ}/\Omega_{\rm H})c}{GM_{\rm BH}^2} \\
    &- \frac{2\chi_{\rm BH}(\dot{M}c^2E_{\rm ISCO} - L_{\rm BZ})}{M_{\rm BH}c^2}.
\end{split}
\end{equation}

By combing Equations (\ref{equ:BZeffect}), (\ref{equ:dM_BHdt}) and (\ref{equ:dchi_BHdt}), the time-evolving BZ power can thus be calculated. The observed $\gamma$-ray/X-ray light curve caused by the fall-back accretion is connected to the BZ power via the $\gamma$-ray/X-ray radiation efficiency $\eta_{(\gamma\rm ,X)}$ and the jet beaming factor $f_{\rm b}$, i.e.,
\begin{equation}
    \eta_{(\gamma\rm ,X)}L_{\rm BZ} = f_{\rm b}L_{(\gamma\rm ,X)}.
\end{equation}

\subsubsection{Jet Afterglow Emissions}

In order to calculate the afterglow light curves, we adopt the Gaussian structured jet model \citep[e.g.,][]{zhang2002} which was favored by the observations of GRB\,170817A afterglow \citep[e.g.,][]{lamb2018,lazzati2018,mooley2018,troja2018,xie2018}, i.e.,
\begin{equation}
    E(\theta) = E_0\exp\left(-\frac{\theta^2}{2\theta_{\rm c}^2}\right),
\end{equation} 
where $E_0$ is the on-axis equivalent isotropic energy and $\theta_{\rm c}$ is the characteristic core angle. The spectra of the standard synchrotron emission from relativistic electrons are employed following \cite{sari1998,kumar2015} and \cite{zhang2018}. For more details of the afterglow modeling we applied to calculate the sGRB light curves along the line of sight, see Appendix C in \cite{zhu2021kilonova}. We constrain the afterglow parameters, including $E_{\rm 0}$, $\theta_{\rm c}$, viewing angle $\theta_{\rm v}$ with respect to the moving direction of the jet, circumburst number density $n$, power-law index of the electron distribution $p$, ānd fraction of shock energy carried by magnetic fields $\varepsilon_B$, to fit the multi-band light curves of GRB\,211211A. The fraction of shock energy carried by electrons is set to its typical value of $\varepsilon_e = 0.1$.

\subsubsection{Ejecta Mass}

After NSBH mergers, a fraction of neutron-rich matter (i.e., an unbound dynamical ejecta) is tidally ejected while an accretion disk is formed around the remnant BH. The total remnant mass outside the remnant and the dynamical ejecta mass are dependent on the NSBH system parameters, including the BH mass $M_{\rm BH}$, the dimensionless spin parameter projected onto the orientation of orbital angular
momentum $\chi_{\rm BH}$, the NS mass $M_{\rm NS}$, and the NS equation of state (EoS), which can be calculated based on an empirical fitting formula, i.e.,
\begin{equation}
\label{equ:FittingFormula}
    \frac{M_{\rm fit}}{M^{\rm b}_{\rm NS}} = \left[\max\left(a_1\frac{1 - 2C_{\rm NS}}{\eta^{1 / 3}} - a_2 \widetilde{R}_{\rm {ISCO}}\frac{C_{\rm NS}}{\eta} + a_3 , 0 \right)\right]^ {a_4},
\end{equation}
where $ M^{\rm b}_{\rm NS}$ is the baryonic mass of the NS, $C_{\rm NS}$ is the compactness of the NS determined by the NS EoS, $\eta = Q / (1 + Q) ^ 2$, and $Q = M_{\rm BH} / M_{\rm NS}$ is the mass ratio between the primary BH mass and the secondary NS mass. For the fitting formula of the total remnant mass $M_{\rm tot,fit}$ (the dynamical ejecta mass $M_{\rm d,fit}$), the parameters in Equation (\ref{equ:FittingFormula}) are $a_1 = 0.406$, $a_2 = 0.139$, $a_3 = 0.255$, and $a_4 = 1.761$ ($a_1 = 0.218$, $a_2 = 0.028$, $a_3 = -0.122$, and $a_4 = 1.358$) obtained from \cite{foucart2018} \citep{zhu2020}. Since the fitting formulas of the total remnant mass and the dynamical ejecta mass are obtained with independent simulation data, one needs to set an upper limit on the maximum fraction of dynamical ejecta mass to the total remnant mass, i.e., $M_{\rm d,max} \approx f_{\rm max}M_{\rm total,fit}$. We set $f_{\rm max} \approx 0.5$ based on simulation results from \cite{kyutoku2015}. Therefore, the final empirical mass of the dynamical ejecta is $M_{\rm d} \approx \min(M_{\rm d,fit} , f_{\rm max}M_{\rm total,fit})$.

We consider two ejecta components for NSBH kilonova model, i.e., the wind ejecta from the disk around the remnant BH and the dynamical ejecta caused by tidal forces. The wind ejecta mass can be estimated as a constant fraction of the disk mass, i.e., $M_{\rm w} \approx \xi_{\rm w}M_{\rm disk}$, where $\xi_{\rm w}\approx0.2$ \citep{fernandez2015,just2015,siegel2017}.

Numerical simulations revealed that the dynamical ejecta from NSBH mergers are highly anisotropic and distributed in the equatorial plane \citep{kyutoku2015,kawaguchi2016,darbha2021}. \cite{zhu2020} constructed a viewing-angle-dependent model for NSBH kilonovae and found that the wind ejecta can be covered by the dynamical ejecta for a large $\theta_{\rm v}$ condition. However, for the case of GRB\,211211A observed in the on-axis or near-on-axis view (i.e., $\theta_{\rm v}\sim0^\circ$), one can simultaneously see two components. Hereafter, in order to reduce computational complexity, we used a simplified model based on \cite{zhu2020} to separately consider the emissions from the wind ejecta and the dynamical ejecta.

Assuming that the wind ejecta has an isotropic density profile and a homologous expansion, we adopt the common analytic solution derived by \cite{arnett1982} and \cite{chatzopoulos2012} to calculate the bolometric luminosity of the wind ejecta:
\begin{equation}
    L_{\rm{w}}(t) = e^{-(t'/t_{\rm{w,diff}})^2}\int_0^t2L_{\rm{w, in}}(t')\frac{t'}{t_{\rm{w,diff}}}e^{(t'/t_{\rm{w,diff}})^2}\frac{dt'}{t_{\rm{w,diff}}},
\end{equation}
where $t_{\rm{w,diff}} = (2\kappa_{\rm{w}}M_{\rm{w}}/\beta{v_{\rm{w}}c})^{1/2}$ is the photon diffusion timescale of the wind ejecta, $\kappa_{\rm{w}}$ is the grey opacity, and $\beta = 13.8$ is the dimensionless constant. $L_{\rm{w,in}}(t) = \epsilon_{Y_e}\epsilon_{\rm th}\dot{\epsilon}(t)M_{\rm{w}}$ represents the injection heating rate from the radioactive decay of $r$-process nucleus, where $\epsilon_{Y_e} = 0.5 + 2.5[1+e^{4(t/{\rm day} - 1)}]^{-1}$ if $Y_e\leq0.25$ ($\epsilon_{Y_e} = 1$ otherwise) is an electron-fraction-dependent term which takes into account extremely neutron-rich ejecta with a decay half-life of a few hours \citep{perego2017}, $\epsilon_{\rm{th}}\approx0.5$ is the efficiency of thermalization \citep{metzger2010}, and $\dot{\epsilon}(t) = \dot{\epsilon}_0(t/{\rm day})^{-\alpha}$ is the specific energy injection rate due to radioactive decay with $\dot{\epsilon}_0 \approx 1.58\times10^{10}\,{\rm erg}\,{\rm g}^{-1}\,{\rm s}^{-1}$ and $\alpha \approx 1.3$ \citep{korobkin2012}. 

In order to calculate the monochromatic light curves of the wind ejecta, we define an photosphere temperature as
\begin{equation}
    T_{\rm{w,phot}}(t) = \max\left[\left(\frac{L_{\rm{w}}(t)}{4\pi\sigma_{\rm SB}v_{\rm{w}}^2t^2}\right)^{1/4} , T_{\rm La}\right].
\end{equation}
where $\sigma_{\rm SB}$ is the Stefan-Boltzmann constant, $v_{\rm w}\approx0.067\,c$ is assumed to be the wind ejecta velocity \citep[e.g.,][]{just2015,siegel2017,perego2017}, and $T_{\rm La}\approx1000\,{\rm K}$ is the first ionisation temperature of lanthanides \citep{barnes2013}. The photosphere radius can be written as
\begin{equation}
    R_{\rm{w,phot}}(t) = \left\{\begin{array}{ll}v_{\rm{w}}t, & {~\rm for~}T_{\rm{w,phot}} > T_{\rm La}, \\
    \left(\frac{L_{\rm{w}}(t)}{4\pi{\sigma_{\rm{SB}}}T_{\rm{La}}^4}\right)^{1/2}, & {~\rm for~}T_{\rm{w,phot}} = T_{\rm La},
\end{array}\right.
\end{equation}
Then, the flux density contributed from the emission of the wind ejecta is given by
\begin{equation}
    F_{\nu,{\rm{w}}} = \frac{8\pi^2R_{\rm{w,phot}}^2}{h^2c^2}\frac{h^3\nu^3}{\exp({h\nu/k_{\rm B}T_{\rm{w,phot}}} - 1)}\frac{1}{4\pi D_{\rm L}^2},
\end{equation}
where $h$, $k_{\rm B}$ and $\nu$ represent the Planck constant, the Boltzmann constant and frequency, respectively.

Based on the simulations of NSBH mergers \citep{kyutoku2015}, the mass distribution of dynamical ejecta is highly anisotropic, with the mass mainly distributed around the equatorial plane and shaped like a crescent. The dynamical ejecta is typically concentrated around the orbital plane with a half opening angle in the latitudinal direction of $\theta_{\rm d}\approx15^\circ$ and often sweeps out only a half of the plane, i.e., an opening angle in the longitudinal direction of $\varphi_{\rm d}\approx180^\circ$. Since the dynamical ejecta is geometrically thin in the latitudinal direction, the photons would be always diffused from the latitudinal edge. Due to $dM_{\rm d}/dv\approx{\rm const}$ between the radial velocity range of $v_{\rm d,min}<v<v_{\rm d,max}$ based on the numerical relativity simulations \citep{kyutoku2015}, the bolometric luminosity of the dynamical ejecta can be obtained from \cite{kawaguchi2016}, i.e.,
\begin{equation}
    L_{\rm{d}}(t)\approx(1+\theta_{\rm{d}})\epsilon_{\rm th}\dot{\epsilon}_0M_{\rm{d}}\times\left\{\begin{array}{ll}
        \frac{t}{t_{\rm c}}\left(\frac{t}{{\rm day}}\right)^{-\alpha}, & {\rm for~}t<t_{\rm{c}}, \\
        \left(\frac{t}{{\rm day}}\right)^{-\alpha}, & {\rm for~}t>t_{\rm{c}},
    \end{array}\right.
\end{equation}
where $t_{\rm c} = [\kappa_{\rm d}\theta_{\rm d}M_{\rm d} / 2c\varphi_{\rm d}(v_{\rm d,max} - v_{\rm d,min})]^{1/2}$ is defined as the critical diffuse timescale that all ejecta can be seen. We set the minimum velocity $v_{\rm d,min}\approx0.1\,c$ following the simulations of \cite{kyutoku2015} while the maximum velocity can be estimated as $v_{\rm d,max} = \sqrt{3v_{\rm d,rms}^2 - 3v_{\rm d,min}^2/4} - v_{\rm d,min} / 2$ where $v_{\rm d,rms} = (-0.441Q^{-0.224}+0.539)\,c$ is the root-mean-square velocity of the dynamical ejecta obtained from \cite{zhu2020}.

Because $dL_{\rm d}/2$ is released over an area of $\varphi_{\rm d}vt^2dv$, one can derive the photosphere temperature of each velocity at a given time:
\begin{equation}
    T_{\rm{d,phot}}(v , t) \approx \left[\frac{L_{\rm d}(t)}{2\sigma_{\rm SB}\varphi_{\rm d}v(v_{\rm d,max} - v_{\rm d,min})t^2}\right]^{1/4}.
\end{equation}
The total flux density from the dynamical ejecta can be expressed as
\begin{equation}
    F_{\nu,{\rm{d}}} \approx \int^{v_{\rm d,max}}_{v_{\rm d,min}}\frac{4\pi\varphi_{\rm d}vt^2}{h^2c^2}\frac{h^3\nu^3}{\exp(h\nu/k_{\rm B}T_{\rm d,phot} - 1)}\frac{1}{4\pi D_{\rm L}^2}dv.
\end{equation}

\subsection{Origins of GRB\,211211A and Associated Kilonova}

\begin{deluxetable}{ccccc}[t] \label{tab:FittingParameters}
\tablecaption{Priors and Results for Fitting Parameters}
\tablecolumns{3}
\tablewidth{0pt}
\tablehead{
\colhead{Parameter} &
\colhead{Prior} &
\colhead{Min} &
\colhead{Max} &
\colhead{Result}
}
\startdata
$M_{\rm BH}/M_\odot$ & Flat & 2.22 & 15 & {$8.21^{+0.77}_{-0.75}$} \\
$\chi_{\rm BH}$ & Flat & 0 & 0.997 & {$0.62^{+0.06}_{-0.07}$} \\
$M_{\rm NS}/M_\odot$ & {{Gaussian}} & {{1.0}} & 2.22 & {$1.23^{+0.06}_{-0.07}$} \\
$\kappa_{\rm w}/{\rm cm}^2{\rm g}^{-1}$ & Flat & 0.5 & 5 & {$0.56^{+0.13}_{-0.05}$}\\
$\kappa_{\rm d}/{\rm cm}^2{\rm g}^{-1}$ & Flat & 10 & 100 & {$11.0^{+1.4}_{-0.8}$} \\
$\dot{M}_{\rm p}/M_\odot\,{\rm s}^{-1}$ & Log-flat & $10^{-10}$ & $1$ & {$2.19^{+0.44}_{-0.37}\times10^{-5}$} \\
$t_0/{\rm s}$ & Log-flat & 10 & 12.7 & $11.2^{+0.0}_{-0.0}$ \\
$t_{\rm p}/{\rm s}$ & Log-flat & 12.7 & 40 & {$30.2^{+0.0}_{-0.0}$} \\
$t_{\rm b}/{\rm s}$ & Log-flat & 100 & 1000 & $234^{+0}_{-0}$\\
$s$ & Flat & 0 & 15 & {$7.33^{+0.19}_{-0.19}$}\\
$\eta_\gamma$ & Log-flat & $10^{-2}$ & 1 & $0.29^{+0.00}_{-0.00}$ \\
$E_0/{\rm erg}$ & Log-flat & $10^{50}$ & $10^{53}$ & {$5.1^{+2.6}_{-1.9}\times10^{52}$} \\
$\theta_{\rm c}/{\rm rad}$ & Flat & 0 & 0.2 & $0.03^{+0.00}_{-0.00}$\\
$\theta_{\rm v}/{\rm rad}$ & Flat & 0 & 0.5 & $0.07^{+0.01}_{-0.01}$ \\
$p$ & Flat & 2 & 3 & {$2.01^{+0.01}_{-0.01}$} \\
$n/{\rm g}\,{\rm cm}^{-3}$ & Log-flat & $10^{-6}$ & $2$ & {$0.41^{+0.77}_{-0.30}$}\\
$\epsilon_B$ & Log-flat & $10^{-5}$ & 1 & {$1.3^{+1.8}_{-0.6}\times 10^{-3}$}
\enddata
\tablecomments{{{The prior of $M_{\rm NS}$ is adopted as a Gaussian distribution, i.e., $\mathcal{N}(M_{\rm NS}/M_\odot)\sim(\mu = 1.3 , \sigma = 0.11)$ , consistent with the observed mass distribution of NSs in Galactic BNS systems \citep{lattimer2012}. }}  }
\end{deluxetable}

\begin{figure*}
    \centering
	\includegraphics[width = 0.49\linewidth , trim = 60 30 80 50, clip]{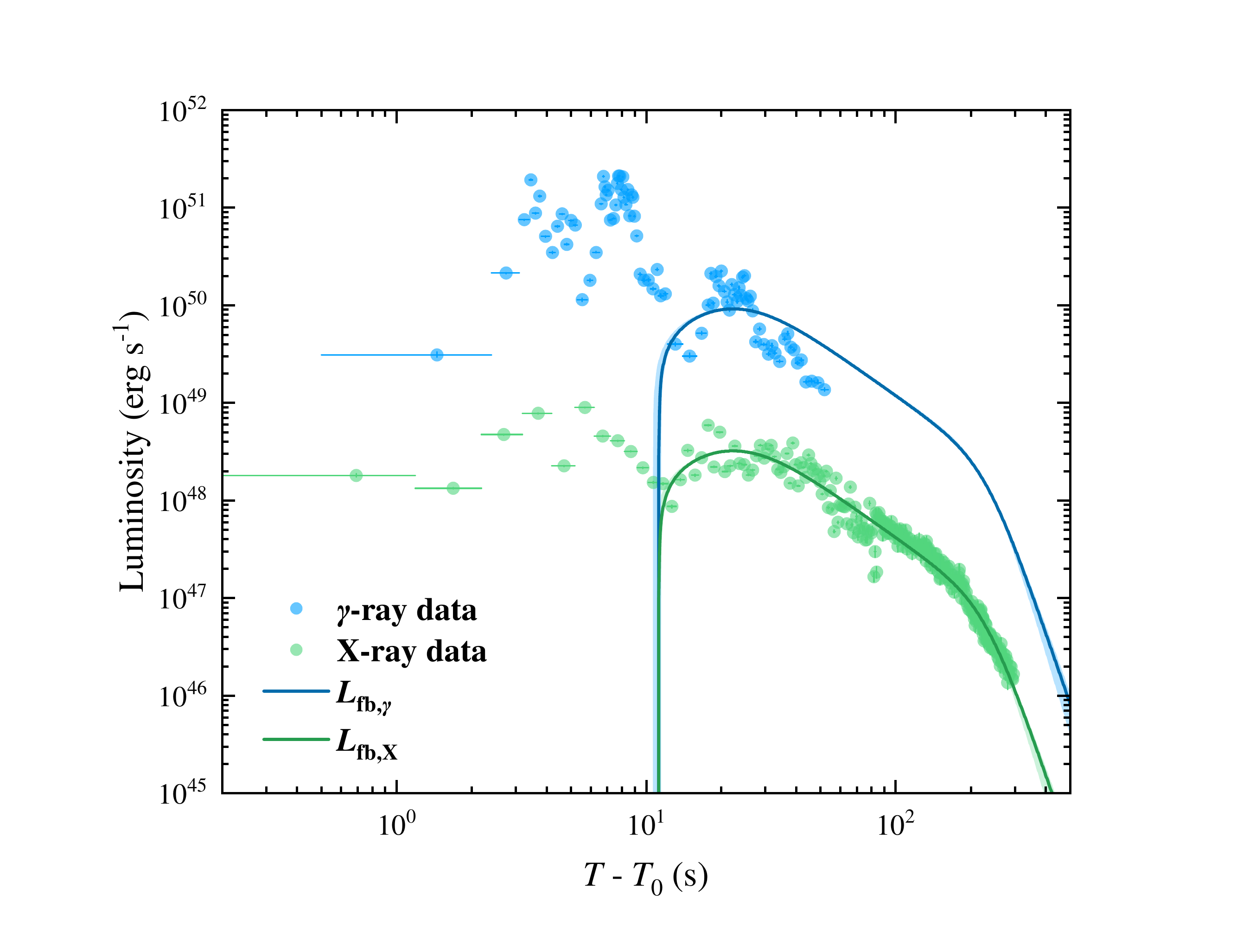}
	\includegraphics[width = 0.49\linewidth , trim = 60 30 80 50, clip]{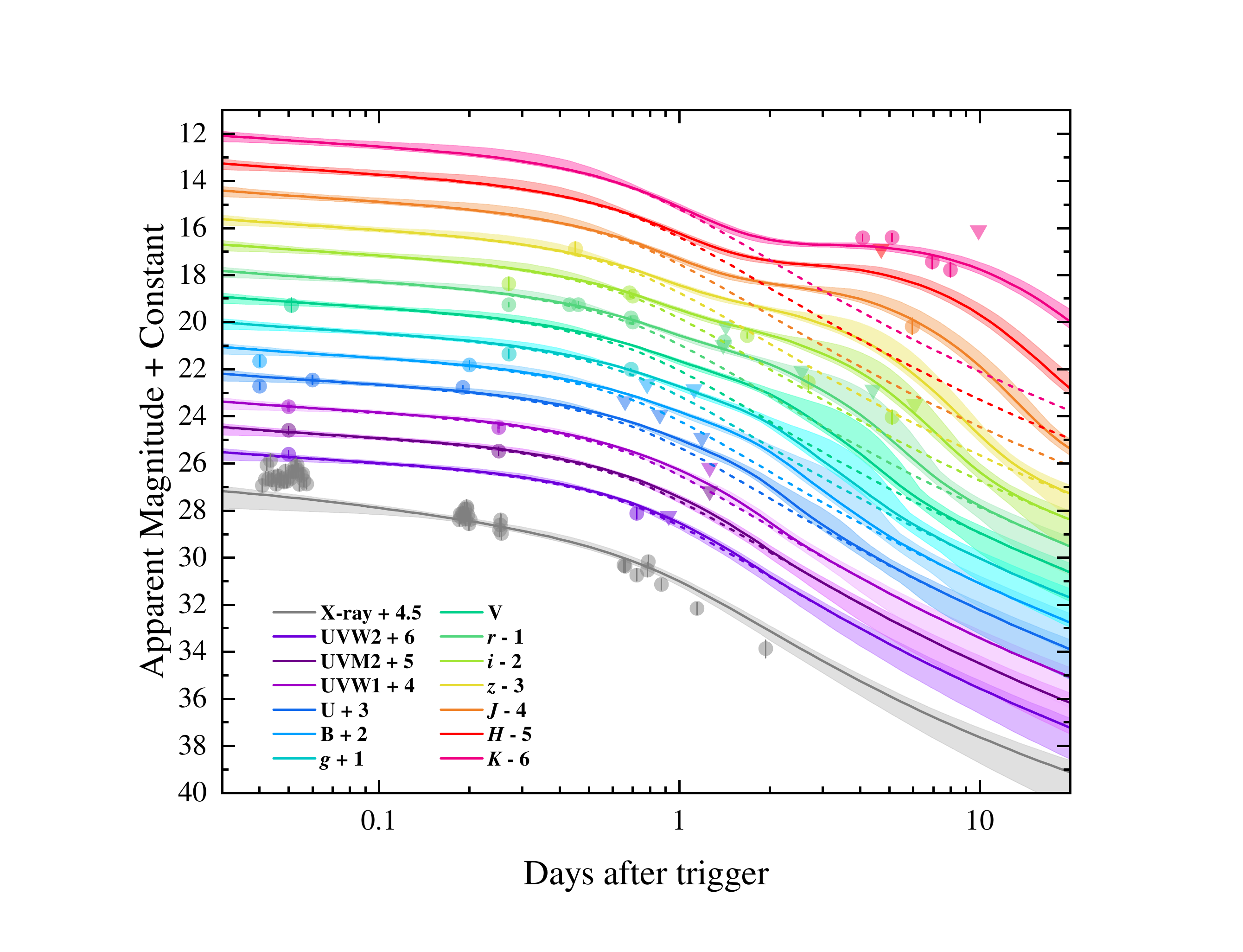}
    \caption{{Left panel:} $\gamma$-ray \citep[blue points from][]{yang2022} and X-ray (green points from BAT and XRT) light curves of GRB\,211211A. Blue and green lines are the fittings of the EEs using the fall-back accretion model. {Right panel:} the detections (circle points) and upper limits (inverted triangle points) of the multi-band data for GRB\,211211A afterglow and associated kilonova emissions. The solid lines and shaded areas represent the best-fittings and 90\% credible intervals for the multi-band data, while the contributions from the afterglow emissions are marked as the dashed lines. }
    \label{fig:Fitting}
\end{figure*}

For simplicity, we directly use the final BH mass $M_{\rm BH,f}$ and final dimensional aligned-spin $\chi_{\rm BH,f}$ after NSBH mergers based on {{Equation (3) and Equation (5) from \cite{deng2020}}} as functions of the initial NSBH system parameters, i.e., $M_{\rm BH}$, $M_{\rm NS}$, $C_{\rm NS}$ and $\chi_{\rm BH}$, to determine the BZ power.The observed mass distribution of Galactic BNS systems \citep{lattimer2012} were inferred to be a Gaussian distribution, i.e., $\mathcal{N}(M_{\rm NS}/M_\odot)\sim(\mu = 1.3 , \sigma = 0.11)$. The observations of GW200105 and GW200115 showed that their NS masses are $\sim1.9\,M_\odot$ and $\sim1.5\,M_\odot$ \citep{abbott2021nsbh}, plausibly more massive than the mass distribution of Galactic BNS systems. Furthermore, some population synthesis simulations predicted that NSBH mergers might usually contain more massive NSs compared to those in BNS mergers \citep[e.g.,][]{giacobbo2018,broekgaarden2021b}. However, due to currently limited observations for NSBH GWs that could hardly represent a complete mass distribution of NSs in NSBH mergers, we use the observed mass distribution of NSs in Galactic BNS systems as the prior of NS mass. Here, an EoS of AP4 \citep{akmal1997} is adopted since it is one of the most likely EoSs constrained by GW170817 \citep{abbott2018gw170817}. In our calculations, we set $\eta_{\rm X} = 0.01$ and $f_{\rm b} = 0.01$ while $\eta_\gamma$ is assumed to be a fitting parameter. 

In theory, NSBH kilonovae were thought to be optically dim, but infrared bright compared with BNS kilonovae \citep[e.g.,][]{kasen2017,kawaguchi2020,zhu2020}, because the NSBH merger probably produces a large number of lanthanide-rich dynamical ejecta with opacity $\kappa_{\rm d}\sim10-100\,{\rm cm}^2\,{\rm g}^{-1}$. The disk wind ejecta is hardly to be lanthanide-poor due to the lack of shock heating and neutrino irradiation during or shortly after the merger \citep[e.g.,][]{just2015,fernandez2015}. However, \cite{fujibayashi2020} and \cite{kyutoku2020} recently found that the wind ejecta can still be lanthanide-poor if the viscous coefficient is not extremely high. Given the uncertainty of the wind ejecta, we adopt a wide prior distribution for the wind ejecta opacity, which is $\kappa_{\rm w}\sim0.5-5\,{\rm cm}^2\,{\rm g}^{-1}$.

\begin{figure*}
    \centering
	\includegraphics[width = 1\linewidth]{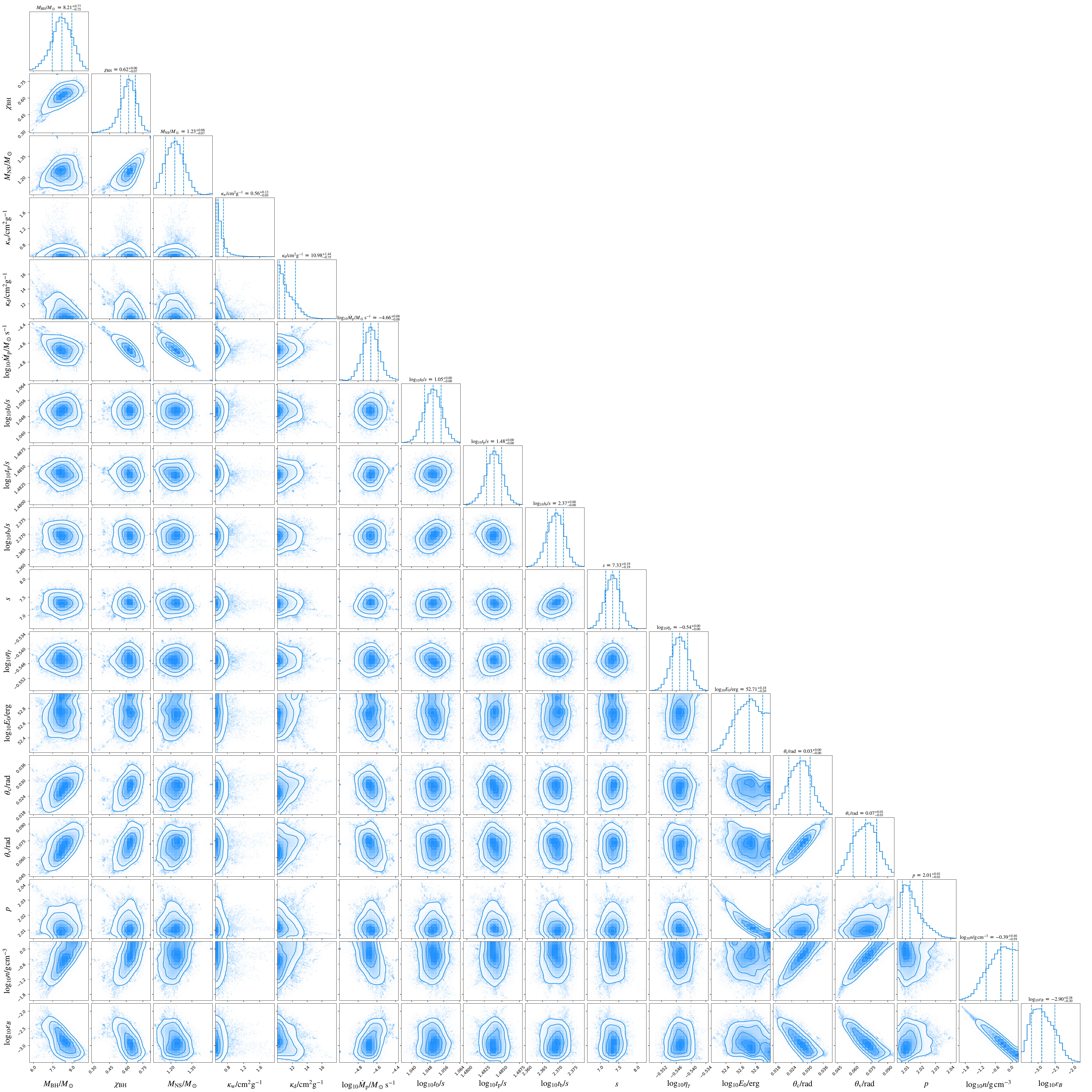}
    \caption{{{Posteriors of the fitting parameters. Medians and 1$\sigma$ credible intervals are labeled.}}}
    \label{fig:Corner}
\end{figure*}

There are 17 free parameters summarized in Table \ref{tab:FittingParameters}. The Markov Chain Monte Carlo method with the \texttt{emcee} package \citep{foremanmackey2013} is adopted to simultaneously fit the data of $\gamma$-ray/X-ray EE, afterglow, and kilonova emissions of GRB\,211211A. We summarize the total 17 free parameters and their fitting results with $1\sigma$ credible intervals in Table \ref{tab:FittingParameters}, {{while the posteriors of these fitting parameters are shown in Figure \ref{fig:Corner}.}} {{Except $\kappa_{\rm w}$ and $\kappa_{\rm d}$, these fitting parameters are convergent. The lower value of $\kappa_{\rm w}$ indicates that the wind ejecta of NSBH mergers could be lanthanide-poor.}} The best-fit light curves of $\gamma$-ray/X-ray EE, afterglow, and kilonova emissions are shown in Figure \ref{fig:Fitting}. 

Our fitting results reveal that GRB\,211211A could be a merger between a {$\sim1.23^{+0.06}_{-0.07}\,M_\odot$} NS and a {$\sim8.21^{+0.77}_{-0.75}\,M_\odot$} BH with an aligned-spin of {$\chi_{\rm{BH}}\sim0.62^{+0.06}_{-0.07}$}. The merger would produce $\sim0.005-0.03\,M_\odot$ lanthanide-poor wind ejecta and $\sim0.015-0.025\,M_\odot$ lanthanide-rich dynamical ejecta. In Section \ref{sec:EventRate}, we suspected that GRB\,211211A can be originated from NS-first-born NSBH mergers based on the estimations of the event rate density. \cite{hu2022} found that for a NS-first-born NSBH binary system, the companion helium star would be tidally spun up efficiently by the NS, and would thus finally form a fast-spinning BH whose aligned-spin is always $\chi_{\rm BH}\gtrsim0.8$. Thus, the mass and spin of the BH component for GRB\,211211A are essentially consistent with those of NS-first-born NSBH merger predicted by \cite{hu2022}. 

Our fitting results show that the fall-back accretion for the interpretation of the EE starts at $t_0 = 11.2\,{\rm s}$, peaks at $t_{\rm p} = 30.2\,{\rm s}$ and breaks around $t_{\rm b} = 234\,{\rm s}$ in the rest frame. The peak fall-back accretion rate is $\dot{M}_{\rm p}\sim2\times10^{-5}\,M_\odot\,{\rm s}^{-1}$. By Equation (\ref{equ:fb}), one can estimate the fall-back mass as $M_{\rm fb}\simeq\int^{t_{\rm b}}_{t_0}\dot{M}dt\approx1\times10^{-3}\,M_\odot$. The start time and peak accretion rate are consistent with the predictions by \cite{desai2019}. Furthermore, \cite{desai2019} predicted that rebrightening EE caused by the fall-back accretion of $r$-process heating materials may only happen if the remnant BH has a mass of $\gtrsim6-8\,M_\odot$. Based on the fitting results, GRB\,211211A would finally form a $\sim10\,M_\odot$ BH after the NSBH merger, so that our interpretations for the origin of GRB\,211211A are self-consistent.

{{There are some uncertainties in our fitting results. Our results reveal that the mass ratio between the BH and NS is $\sim7$ for GRB\,211211A. Due to present limited simulations for NS-first-born NSBH mergers, it is not sure if NSBH systems with such high mass ratio are common in the universe. More detailed simulations for the populations of NS-first-born NSBH mergers based on population synthesis and detailed binary evolution are suggested in the future. We also find $\sim0.005-0.03\,M_\odot$ lanthanide-poor wind ejecta is needed in order to explain the observations of the kilonova emission associated with GRB\,211211A. A large amount of lanthanide-poor wind ejecta typically do not expect to be produced after NSBH mergers \citep[e.g.,][]{just2015,fernandez2015}, although \cite{fujibayashi2020} and \cite{kyutoku2020} found that NSBH mergers still can lead to considerable lanthanide-poor wind ejecta if the viscous coefficient is not extremely high. Future multimessenger observations between NSBH GWs and associated kilonova emissions will help us constrain the mass fraction of $r$-process elements for the wind ejecta from NSBH mergers.}} 

\section{Discussions and Conclusion}

In this {\em Letter}, we collect three unique merger-origin bursts, i.e., GRB\,060614, GRB\,211211, and GRB\,211227A, to study their observed properties and explore possible origins. Both MEs and EEs of these bursts are long-duration, which fall into the distribution of lGRB populations. When the redshift information is considered, we find their WEs of these three bursts still behave as normal lGRBs, but their MEs lie on the sGRB track of the Amati relation. These similar observed properties are characterized differently than those of classical collapsar-origin lGRBs and merger-origin sGRBs with/without EEs, indicating a unique origin for these three bursts. Their X-ray MEs and EEs show unambiguous fall-back accretion signatures, decreasing as $\propto{t}^{-5/3}$, which extend the burst durations. The EEs might result from the fall-back of $r$-process heating materials, predicted to occur after NSBH mergers. The beaming-corrected local event rate density of these merger-origin lGRBs is estimated to be $\mathcal{R}_{0,\rm b}\sim2.4^{+2.3}_{-1.3}(f_{\rm b}/0.01)\,{\rm Gpc}^{-3}{\rm yr}^{-1}$. This local event rate density is much lower than that of BNS and NSBH mergers in the universe but consistent with the local event rate density of NS-first-born mergers.

Our detailed analysis on the EE using the fall-back accretion model, afterglow and kilonova of the recently high-impact event GRB\,211211A reveals it could be a merger between a {$\sim1.23^{+0.06}_{-0.07}\,M_\odot$} NS and a {$\sim8.21^{+0.77}_{-0.75}\,M_\odot$} BH with a dimensionless aligned-spin parameter of {$\chi_{\rm{BH}}\sim0.62^{+0.06}_{-0.07}$}, supporting an NS-first-born NSBH formation channel. We find that the fall-back accretion for the interpretation of the EE starts at $t_0 = 11.2\,{\rm s}$ and peaks at $t_{\rm p} = 30.2\,{\rm s}$ with a peak accretion rate of $\dot{M}_{\rm p} \sim 2\times10^{-5}\,M_\odot\,{\rm s}^{-1}$. The fall-back mass is $M_{\rm fb} \sim 1\times10^{-3}\,M_\odot$. The start time and peak accretion rate are consistent with the fall-back accretion of $r$-process heating materials predicted by \citep{desai2019}. {{After the completion of this {\em Letter}, we notice that \cite{Meng2022} also showed GRB\,211211A can originate from a NSBH system in the photosphere emission model whose long duration is from the duration stretching effect of the saturated photosphere.}} Furthermore, \cite{yang2015} reported that the kilonova candidate associated with GRB\,060614 had an ejection of $\sim0.1\,M_\odot$ of $r$-process material. {{\cite{yang2015} suggested that}} such significant ejected mass, {{within the possible range of dynamical ejecta of mergers between NSs and BHs with extreme high aligned-spins \citep[e.g.,][]{lovelace2013,kyutoku2015}}}, strongly favored its origin for a NSBH merger rather than a BNS merger. Since NS-first-born NSBH mergers can easily occur tidal disruption while the rest of BH-first-born NSBH mergers mostly contribute to plunging events, NSBH mergers can well interpret the origins of these GRBs. Long-duration burst with rebrightening fall-back accretion signature of $r$-process heating materials after MEs, and bright kilonova emission might be commonly observed features for on-axis NSBH mergers.

Based on the estimated local event rate of merger-origin lGRBs, if they are certainly originated from NS-first-born NSBH mergers, the GW detection rate of NSBH mergers with fast-spinning primary BHs in the GW fourth observing run (O4) and fifth observing run (O5) of LIGO/Virgo/KAGRA Collaboration are $\sim10\,{\rm yr}^{-1}$ and $\sim100\,{\rm yr}^{-1}$ \citep{zhu2021kilonova}, respectively. By assuming that all of associated kilonova emissions can be detected, we estimate the multimessenger detection rate between GWs, GRB and kilonova emissions from NSBH mergers in O4 and O5 are $\sim0.1(f_{\rm b} / 0.01)\,{\rm yr}^{-1}$ and $\sim1(f_{\rm b} / 0.01)\,{\rm yr}^{-1}$, respectively. Thus, the smoking-gun evidence for NSBH merger origin of lGRB and kilonova will likely be verified in O5.

\acknowledgments

{{The authors acknowledge an anonymous referee for useful discussions}}. We also thank {{Jun Yang}}, Bing Zhang, Bin-Bin Zhang, He Gao, Shun-Ke Ai, and Yun-Wei Yu for helpful comments. This work is supported by the National Natural Science Foundation of China (grant Nos. 11773003, 11833003, 12003028, 12103065, 12121003, 12133003, 12192220, 12192221, U1931201, U2038105), the National Basic Research Program of China (grant No. 2014CB845800), the China Manned Spaced Project (CMS-CSST-2021-B11), the Natural Science Foundation of Universities in Anhui Province (grant No. KJ2021A0106) and the National Key Research and Development Programs of China (2018YFA0404204).

\software{\texttt{Python}, \url{https://www.python.org}; \texttt{HEASoft} \citep{heasoft2014}, \url{http://heasarc.gsfc.nasa.gov/ftools}; \texttt{scikit-learn}, \url{https://scikit-learn.org/stable/index.html}; \texttt{emcee} \citep{foremanmackey2013}};
{{\texttt{corner} \citep{corner}}}

\clearpage

\bibliography{NSBH}{}
\bibliographystyle{aasjournal}

\end{document}